\documentstyle[12pt,epsfig]{article}
\newcommand{\be}{\begin{equation}}
\newcommand{\ee}{\end{equation}}
\def\la{\mathrel{\mathpalette\fun <}}

\def\fun#1#2{\lower3.6pt\vbox{\baselineskip0pt\lineskip.9pt
\ialign{$\mathsurround=0pt#1\hfil##\hfil$\crcr#2\crcr\sim\crcr}}}

\begin{document}

\title{The low-mass $\sigma$-meson: Is it an eyewitness
of confinement? }
\author{V.V. Anisovich and V.A. Nikonov\\
St.Petersburg Nuclear Physics Institute, Gatchina, 188350,
Russia}
\date{November 29, 1999}
\maketitle

\begin{abstract}
In the framework of the dispersion relation
$N/D$-method we restore the low--energy $\pi\pi$
$(IJ^{PC}=00^{++})$-wave amplitude sewing it with the previously
obtained $K$-matrix solution for the region 450--1900 MeV
[V.V. Anisovich, Yu.D. Prokoshkin and A.V. Sarantsev,
Phys. Lett. {\bf B389}, 388 (1996)].
The restored $N/D$-amplitude has a pole on the second sheet of the
complex-$s$ plane, near the $\pi\pi$ threshold. We discuss the
hypothesis that this low--mass pole, or the low-mass
$\sigma$-meson, corresponds to the dynamical state related to the
confinement forces, that is the eyewitness of confinement.
\end{abstract}

The experimental data on meson spectra accumulated by Crystal
Barrel Collaboration \cite{CrBar}, GAMS \cite{GAMS} and BNL \cite{BNL}
groups provided a good basis for the study of
$(IJ^{PC}=00^{++})$-wave, and
the combined $K$-matrix analysis of the reactions $\pi\pi \to \pi\pi$,
$K\bar K$, $\eta\eta$, $\eta\eta'$, $\pi\pi\pi\pi$ was performed over
the mass range 450--1900 MeV  \cite{APS}.
Then the $K$-matrix analysis has been expanded to the waves
$\frac12 0^{+}$ \cite{AlexSar} and
$10^{+}$ \cite{AKPSS}, thus making
it possible to establish the $q\bar q$ systematics of scalars for
$1^3P_0 q\bar q$ and $2^3P_0 q\bar q$ multiplets.

The $K$-matrix amplitude analysis is a good instrument
to perform the $q\bar
q$ nonet classification of mesons in terms of "bare states", namely,
the states without inclusion of the decay channels (detailed discussion
can be found in Ref. \cite{AAS}). The decay processes cause a
strong mixing of scalar states, for in the transitions $(q\bar q)_1
\to real\; mesons \to (q\bar q)_2$ the
orthogonality  of the coordinate wave functions does
not work. The other important effect caused by the decay is
accumulation of widths of  neighbouring resonances by one of them:
as a result, a broad scalar/isoscalar state appears at the mass
region 1200-1600 MeV.

According to Refs. \cite{AlexSar,AKPSS},
the lightest scalar multiplet $1^3P_0 q\bar q$ of bare states is
located in the region  950--1200 MeV. After the mixing originated from
the decay processes, the bare states of the  $1^3P_0 q\bar q$ nonet
are transformed into a set of resonances: $f_0(980)$, $a_0(980)$,
$f_0(1300)$, $K_0(1000)$ (or $K_0(1400)$) which are the descendants of
the lightest pure  $q\bar q$ states. The scalar/isoscalar resonances
$f_0(1500)$ and $f_0(1750)$ are descendants of the bare
$2^3P_0 q\bar q$ states. At
the same region there is a broad resonance, with the mass
about 1200--1600 MeV, which is the descendant of the
lightest scalar glueball: the gluonium and scalar/isoscalar $q\bar
q$ states are strongly mixed (according to the rules of $1/N$
expansion \cite{t'Hooft}, the transition $gluonium \to q\bar q$
is not suppressed), and the $q\bar q$ component in the broad
state reaches 50\%, while the scalars $f_0(1300)$ and $f_0(1500)$ have
considerable admixtures of gluonium components. In $f_0(980)$ and
$f_0(1750)$, the $s \bar s$ component is predominant.

An important result of Refs. \cite{APS,AKPSS}  is
that the $K$-matrix $00^{++}$-amplitude  has no
pole singularities in the region 500--800 MeV. The
$\pi\pi$-scattering phase $\delta^0_0$ increases smoothly in this energy
region reaching 90$^\circ$ at 800--900 MeV. A straightforward
explanation of such  behaviour of $\delta^0_0$ might be the
presence of a broad resonance, with
mass  about 600--900 MeV and
width $\Gamma \sim 500$ MeV \cite{800P,800E,800A,800Ishida}.
However, according to the $K$-matrix solution \cite{APS,AKPSS},
the $00^{++}$-amplitude does not contain  pole sigularities
on the second sheet of the complex-$M_{\pi\pi}$ plane
inside the interval $450 \leq Re\; M_{\pi\pi}\leq 900 $
MeV: the $K$-matrix amplitude
has a  low-mass pole
only, which is located on the  second sheet either near the
$\pi\pi$ threshold or even below it. In Refs. \cite{APS,AKPSS}, the
presence of this pole was not emphasized, for the left-hand cut,
which plays
an important role in the partial amplitude, was  taken
into account in indirect way only. A proper way for the description of
the low-mass amplitude must be using of the dispersion relation
representation.

In this
paper the dispersion relation $N/D$-amplitude
is reconstructed for the $\pi\pi$ scattering
in the region $M_{\pi\pi}< 1000$ MeV, and this amplitude is sewed
with the $K$-matrix solution of Refs.
\cite{APS,AKPSS}. In the next Section, the
dispersion relation amplitude is found using the method
developed in Ref. \cite{AKMS}: the
$N/D$-amplitude provides a
good description of $\delta^0_0$ up to 900 MeV, thus including the
region $\delta^0_0 \sim 90^\circ$.  At the same time
this amplitude does not contain a pole in the region 500--900 MeV;
instead, the pole is located near the $\pi\pi$ threshold.

We suggest that the low-mass pole in the scalar/isoscalar
wave is related to a fundamental phenomenon at large
distances (in hadronic scale). In Section 2
we argue that the low-mass
pole  corresponds to a white composite particle which is inherent to
subprocesses responsible for the colour confinement forces.

\section{Dispersion relation $N/D$-solution for the
$\pi\pi$-scat\-tering amplitude below 900 MeV}

The pion-pion scattering partial amplitude being
 a function of the invariant energy squared,
$s=M_{\pi\pi}^2$, can be
represented as a ratio $N(s)/D(s)$, where $N(s)$ has  left-hand cut,
which is due to the "forces" (the interactions due to $t$- and
$u$-channel exchanges), while the function $D(s)$ is determined by the
rescatterings in the $s$-channel. $D(s)$ is given by the
dispersion integral along the right-hand cut in the complex-$s$ plane:
\be
A(s)=\frac{N(s)}{D(s)}\; , \;\;\;D(s)=1-\int
\limits_{4\mu^2_\pi}^\infty \frac
{ds'}{\pi} \frac{\rho(s')N(s')}{s'-s+i0}\; .
\ee
Here $\rho(s)$ is the invariant $\pi\pi$ phase space,
$\rho(s)=(16\pi)^{-1}
\sqrt{(s-4\mu^2_{\pi})/s}$. It is supposed in (1) that $D(s) \to 1$ with
$s\to \infty$ and CDD-poles are absent (a detailed presention of the
$N/D$-method can be found in \cite{Chew}).

The $N$-function can be written as an integral along the left-hand
cut as follows:
\be
N(s)=\int
\limits_{-\infty}^{s_L}  \frac{ds'}{\pi}\frac{L(s')}{s'-s}\; ,
\ee
where the value $s_L$ marks the beginning of the
left-hand cut. For example,
for the one-meson exchange diagram $g^2/(m^2 -t)$, the
left-hand cut
starts at $s_L=4\mu_\pi^2-m^2$, and the
$N$-function in this point has a logarithmic singularity; for
the two-pion exchange, $s_L=0$.

Below we work with the amplitude $a(s)$, which is defined as follows:
\be
a(s)= \frac {N(s)}{8\pi \sqrt{s}\left (1-P\int
\limits_{4\mu_\pi^2}^\infty
\frac{ds'}{\pi}\frac{\rho(s')N(s')}{s'-s}\right ) }\; .
\ee

The amplitude $a(s)$  is related to the scattering phase shift:
$a(s)\sqrt{s/4-\mu_\pi^2} = \tan
\delta^0_0$.  In Eq. (3) the threshold singularity is singled out
explicitly, so the function $a(s)$  contains left-hand cut only
as well as the poles corresponding to zeros
of the denominator of the
right-hand side (3): $1=P\int\limits _{4\mu_\pi^2}^\infty
(ds'/\pi)\cdot \rho(s')N(s')/(s'-s) $. The pole of $a(s)$
at $s>4\mu_\pi^2$ corresponds
to the phase shift
value $\delta^0_0 = 90^\circ$. The phase of the $\pi\pi$
scattering  reaches the value $\delta^0_0 = 90^\circ$ at $\sqrt{s}=
M_{90}\simeq
850$ MeV. Because of that, the amplitude $a(s)$ may be represented in
the form:
\be
a(s)=\int\limits_{-\infty}^{s_L}  \frac{ds'}{\pi}\frac{\alpha(s')}{s'-s}+
\frac{C}{s-M^2_{90}}+D.
\ee
For the reconstruction of the low-mass amplitude, the parameters
$D,C,M_{90}$ and $\alpha(s)$ have been determined by fitting to the
experimental data.  In the fit we have used the method, which has been
approved in the analysis of the low-energy nucleon-nucleon amplitudes
\cite{AKMS}.  Namely, the integral in the right-hand side of (4) has
been replaced by the sum
\be
\int\limits_{-\infty}^{s_L}
\frac{ds'}{\pi}\frac{\alpha(s')}{s'-s} \to \sum_{n} \frac{\alpha_n}{s_n
-s}
\ee
with $ -\infty < s_n \leq s_L$.

The description of data within the $N/D$-solution,
which uses six terms in the sum (5), is demonstrated on Fig. 1a.
Parameters of the solution are given at Table 1. The scattering length
in this solution is equal to
$a^0_0=0.22\; \mu_\pi^{-1}$, the Adler zero
is at $ s=0.12 \; \mu_\pi^2$. The
$N/D$-amplitude is sewed with the $K$-matrix
amplitude of Refs. \cite{APS,AKPSS}, and figure 1b demonstrates the
level of the coincidence of the amplitudes  $a(s)$ for both solutions
(the values of $a(s)$ which correspond to the $K$-matrix amplitude are
shown with error bars determined in
\cite{APS,AKPSS}).

The dispersion relation solution has a correct analytic structure at
the region $|s|<1$ GeV$^2$.  The amplitude has no
poles on the first sheet of the complex-$s$ plane;
the left-hand cut of the $N$-function after
the replacement given by Eq. (5) is transformed into a set of poles
on the  negative piece of
the real $s$-axis: six poles of the amplitude (at $s/\mu_{\pi}^2=
-5.2,\; -9.6,\; -10.4,\; -31.6,\; -36.0,\; -40.0$)
represent the left-hand singularity of $N(s)$.
On the second sheet (under the $\pi\pi$-cut) the amplitude has two
poles:  at $s\simeq (4-i14)\mu^2_{\pi}$  and $s\simeq
(70-i34)\mu^2_{\pi}$ (see Fig. 2). The second pole, at
$s=(70-i34)\mu^2_{\pi}$, is located beyond the region under
consideration, $|s|<1$ GeV$^2$
(nevertheless, let us stress that the $K$-matrix amplitude
\cite{APS,AKPSS}
has a set of poles just in the region
of the second pole of the $N/D$-amplitude).
The pole near threshold, at
\be
s\simeq (4-i14)\mu^2_{\pi} \; ,
\ee
is what we discuss.
The $N/D$-amplitude has no poles at $Re \sqrt
s \sim 600-900$ MeV despite the phase shift
$\delta^0_0$ reaches $90^\circ$ here.

The data do not fix the $N/D$-amplitude rigidly. The position of the
low-mass pole can be easily varied in the region
$Re\; s \sim (0 - 4)\mu_{\pi}^2$, and there are simultaneous
variations of
the scattering length in the interval $a^0_0 \sim (0.21  - 0.28 )
\mu^{-1}_\mu $ and Adler zero at $s\sim
(0-1)\mu_{\pi}^2$.

The problem of the low-mass $\sigma$-meson was discussed previously.
In the approaches which take into account the left-hand cut, the
following positions of the pole singularities were found:\\
(i) dispersion relation approach, $s \simeq (0.2-i22.5)
\mu_\pi^2 $ \cite{Basdevant}, \\
(ii) meson exchange models, $s \simeq (3.0-i17.8)\mu_\pi^2 $
\cite{Zinn}, $s \simeq (0.5-i13.2)\mu_\pi^2 $ \cite{Bugg},\\ $s \simeq
(2.9-i11.8)\mu_\pi^2 $ \cite{Speth}, \\
(iii) linear $\sigma$-model,
$s \simeq (2.0-i15.5)\mu_\pi^2 $ \cite{Achasov},\\
which are in an agreement with our result.

However in
Refs. \cite{Markushin,Roos}, the pole singularities were obtained in 
the region of higher masses, at $Re\; s \sim (7-10) \; \mu_\pi^2$.

\section{ Low-mass pole as the eyewitness of confinement}

We believe that the
existence of the low-mass pole in the $00^{++}$-amplitude
is not an occasional event. The creation of a
composite scalar/isoscalar particle with  small mass should be
associated with certain fundamental phenomenon at
separations, which are
large in hadronic scale: such a phenomenon may be the confinement.

The confinement interaction is modelled by the scalar potential
in colour octet state, $V_8(r)$; the potential  increases infinetely
at large distances, $V_8(r)\sim r$ at $r \to \infty$.
The formation of the confinement potential is promptly related to the
creation of a set of $q\bar q$-pairs, or
a $q\bar q$-chain. The examples are provided by the
multihadron production at $\mu^+\mu^-$-annihilation (the transition
$\gamma^* \to q\bar q$) or highly excited meson decay
(transition $M^* \to q\bar q$, see Fig. 3a). In the
decay process of Fig. 3a,
the flying away colour quarks produce a chain of the
$q\bar q$-pairs due to which the colour of the upper quark flees
to the bottom quark. The process of Fig. 3a results in cutting
the self-energy diagram of Fig. 3b. The interaction block
inside the quark loop is resposible for the formation of the
colour potential $V_8(r)$;  this block is shown separately in Fig. 3c.

The increase of $V_8(r)\sim r$ at large $r$ means the existence of a
strong singularity in the momentum representation at small $|\vec q|$:
$V_8(\vec q\; ^2) \sim 1/\vec q\; ^4$. Using notations of Fig. 3c,
one has $-\vec q\; ^2 =s$.
So, the set of
diagrams of  Fig. 3c--type
being responsible for the confinement forces
has a strong singularity near $s=0$.

The infinite increase of $V_8(r)\sim r$ at $r\to \infty$, which is
reproduced in
the singular behaviour  $V_8(q^2) \sim 1/q^4$ at $q^2\to 0$,
represents an idealized picture of the confinement. In this picture
the limit $V_8(q^2) \to 1/q^4$ at $q^2\to 0$ can be interpreted
as an exchange by massless composite
colour-octet particle with a  coupling growing infinetely:
$g^2(q^2)/q^2$ with $ g^2(q^2) \sim 1/q^2$ at small
$q^2$.  This composite colour-octet
particle interacts as a whole system only
at large distances: its compositeness  guarantees the return
to standard QCD at small $r$. At large distances, the
strongly interacting colour-octet massless particle provides colour
neutralization of the quark/gluon systems.

However, the colour screening effects (in particular, encountered in
the production of white particles) cut the infinite growth of
$V_8(r)$ at very large $r$ allowing the behaviour
$V_8(r)\sim r$ at $r <R_{confinement} $ only (for example,
see Ref. \cite{Gribov1}).  It results in a softening   of
 singular behaviour of $V_8(q^2)$ at $q^2\sim 0$. Nevertheless,
the $s$-channel pole singularities in the block of Fig. 3c--type
may survive. Our point is that in this case the singularities
at $s\simeq 0$
reveal themseves not only in the colour-octet state but in
 colour-singlet one as well, i.e. in the $\pi\pi$-scattering
amplitude.

To clarify this point, let us re-write the block of Fig. 3c in terms
of the colour/flavour propagators (t'Hooft--Veneziano diagram).
In Fig. 3d, the t'Hooft--Veneziano diagram which corresponds to the
block of Fig. 3c is shown:  solid line represents the propagation of
colour and  dashed line describes that of flavour.
It is seen that

(i) in the $s$-channel the block contains two open colour lines,
$c=3$ and $\bar c=3$, so one has two colour states in the $s$-channel,
octet and singlet:
$c\otimes \bar c =1+8$.

(ii) the two $s$-channel colour amplitudes, octet and singlet ones,
are not suppressed in the $1/N$-expansion ($N=N_c=N_f$), that
means that the same leading-$N$ subgroups of the diagrams are
responsible for the formation of the octet and singlet amplitudes.
Therefore, the positions
of the pole singularities in both amplitudes are
the same in the leading-$N$ terms. However, the next-to-leading terms
(such as decay of the white state into $\pi\pi$)
discharge the rigid degeneration of the octet and singlet
states.

Summing up, in the leading-$N$ terms the pole structure near $q^2=0$ is
the same both for colour--singlet and colour--octet amplitudes. The
colour--octet amplitude providing the growth of the potential
$V_8(r)\sim r$ at $r\sim R_{confinement} $ can have singularities
near $q^2= 0$. This means that  the
colour--singlet amplitude has similar singulariries.
Because of that, corresponding white
state can be named the eyewitness of confinement.

The existence of states, which may be called the
eyewitnesses of confinement, was discussed in Ref. \cite{Gribov2},
though in Ref. \cite{Gribov2} the scalar quartet $(f_0(980),a_0(980))$
was under discussion.

\section{Conclusion}

We have analysed the structure of
the low-mass $\pi\pi$-amplitude in the
region $M_{\pi\pi} \la 900$ MeV using the dispersion relation
$N/D$-method, which provides us with a possibility to take the
left-hand singularities into consideration. The dispersion relation
$N/D$-amplitude is sewed with that given by the $K$-matrix analysis
performed at $M_{\pi\pi} \sim 450-1950$ MeV  \cite{APS}. Obtained in
this way the $N/D$-amplitude has a pole on the second sheet of the
complex-$s$ plane near the $\pi\pi$ threshold. This pole corresponds to
the existence of the low--energy bound state.

On the basis of the conventional confinement model, we argue that the
confinement interaction can produce
low-mass pole singularities both for the
colour--octet and colour--singlet  states. Then the
colour--singlet scalar state, which reveals itself in the low-mass
$\pi\pi$-amplitude, may be named the eyewitness of
 confinement.

\section*{Acknowledgement}
We thank A.V. Anisovich, Ya.I. Azimov, D.V. Bugg, Yu.L.
Dokshitzer, H.R. Petry, A.V. Sarantsev and V.V. Vereschagin for
fruitful discussions.
The article is supported by the RFBR grant N 98-02-17236.

\bigskip
\begin{table}[h]
\caption{Fitting parameters for the amplitude $a(s)$, Eqs. (4), (5).}
\begin{tabular}{|c|c|c|c|c|c|c|}
\hline
$s_n\;\mu^{-2}_\pi$ & -9.56 & -10.16 & -10.76 & -32 & -36 & -40 \\
\hline
$\alpha_n\;\mu^{-1}_\pi$ & 2.21 & 2.21 & 2.21 & 0.246 & 0.246 & 0.246 \\
\hline
\multicolumn{7}{|c|}{$M_{90}=6.228\;\mu_\pi,
\qquad C=-13.64\;\mu_\pi,\qquad D=0.316\;\mu^{-1}_\pi$} \\
\hline
\end{tabular}
\end{table}

\newpage
\begin{figure}
\centerline{\epsfig{file=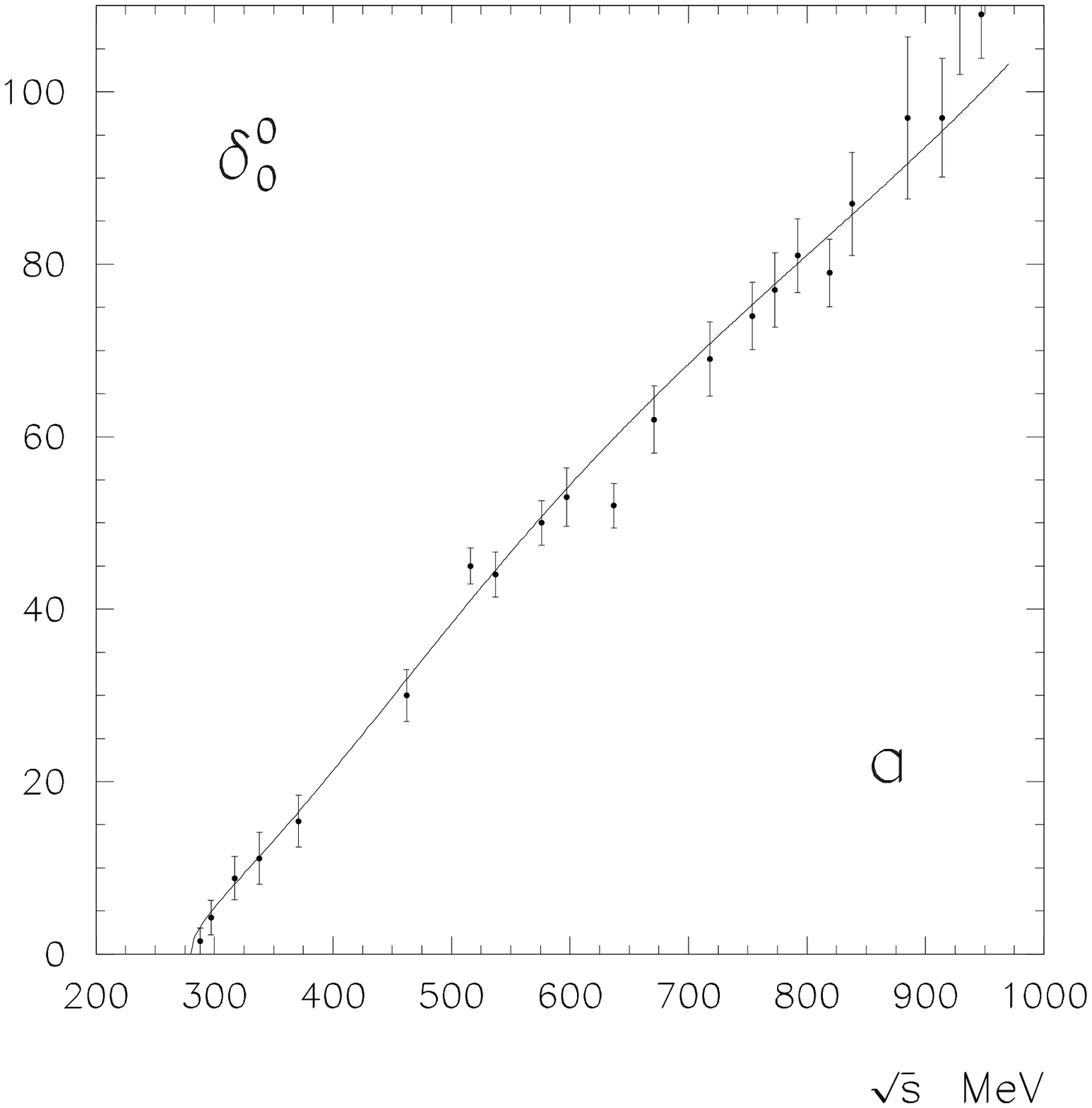,width=7.5cm}\hspace{1cm}
            \epsfig{file=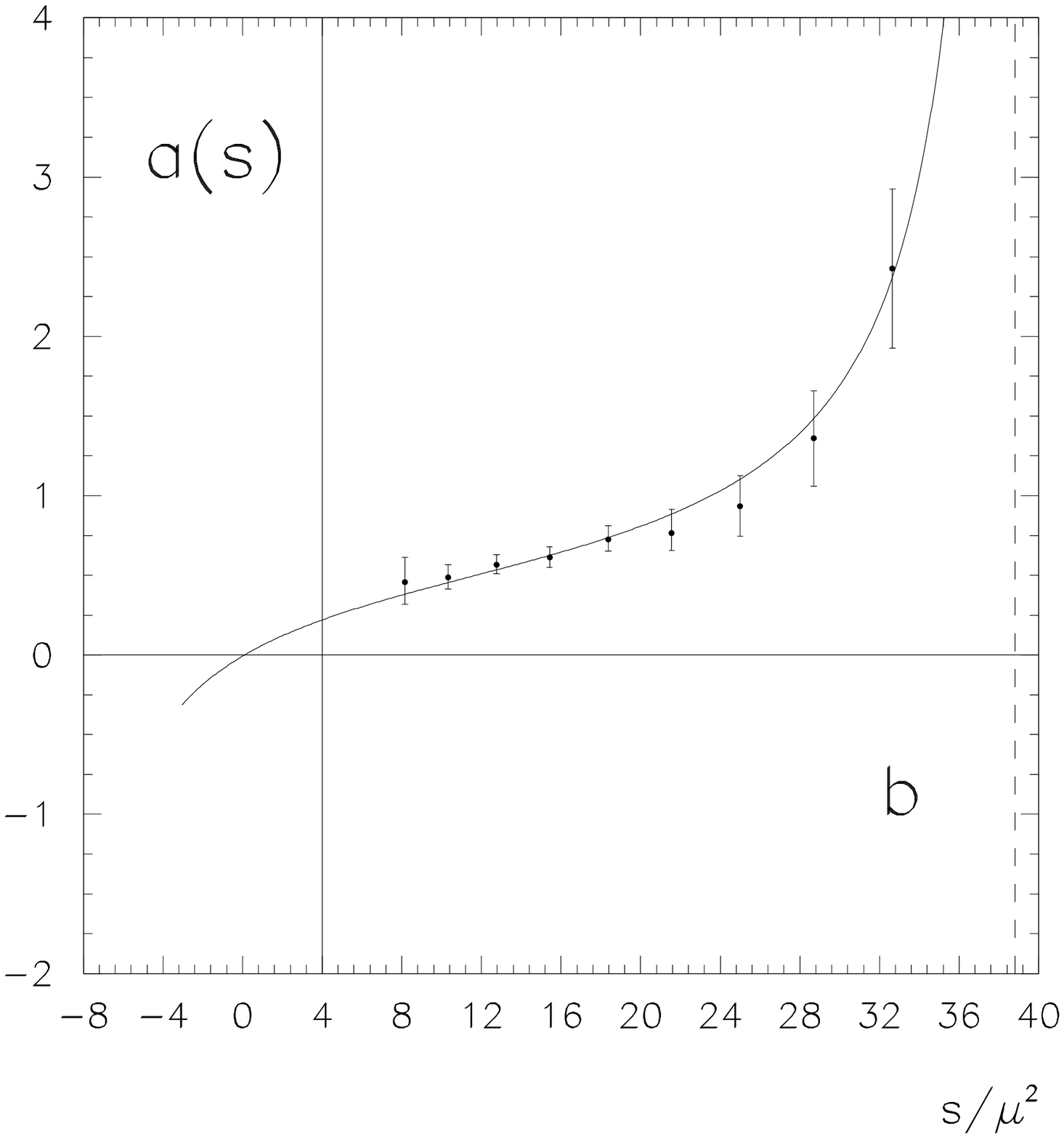,width=7.5cm}}
\caption{a) Fit to the data on $\delta^0_0$ by using the
$N/D$-amplitude.  b) Amplitude $a(s)$ in the $N/D$--solution
(solid curve) and the $K$-matrix approach [4,6] (points with
error bars). }
\end{figure}

\begin{figure}
\vspace{3cm}
\centerline{\epsfig{file=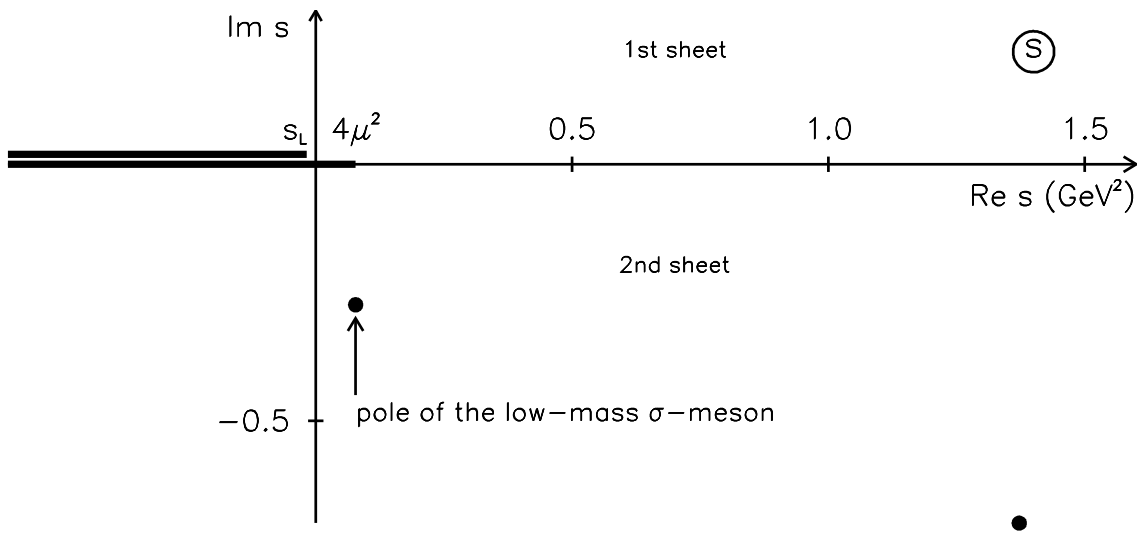,width=15cm}}
\caption{Complex-s plane and singularities of the $N/D$-amplitude}
\end{figure}

\newpage

\begin{figure}
\centerline{\epsfig{file=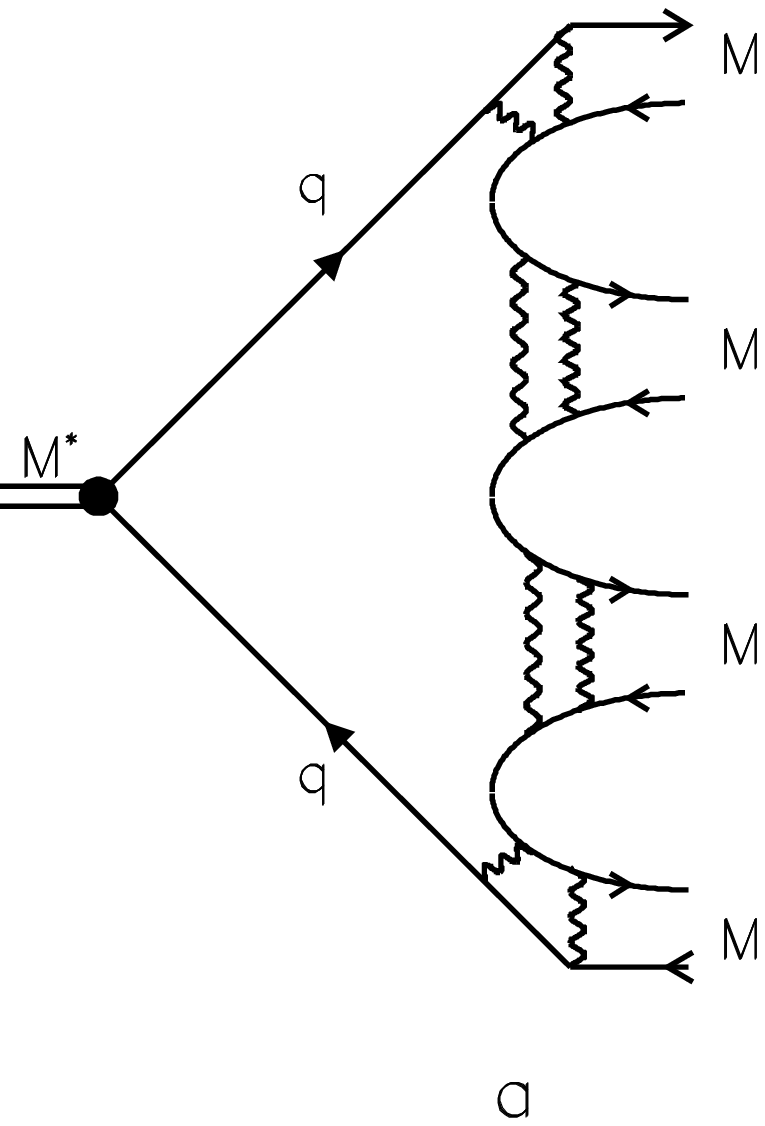,height=8cm}\hspace{1cm}
            \epsfig{file=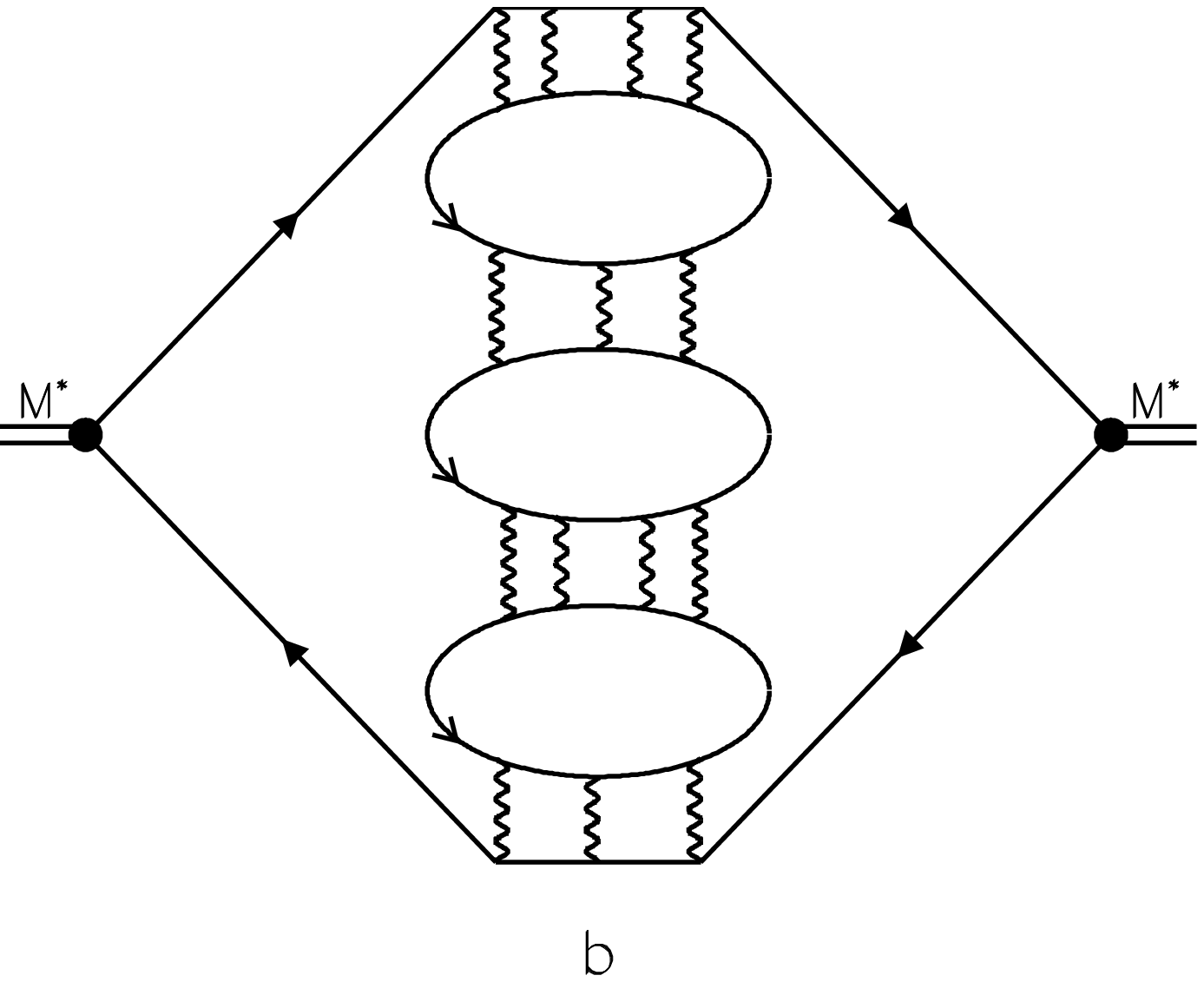,height=8cm}}
\vspace{2cm}
\centerline{\epsfig{file=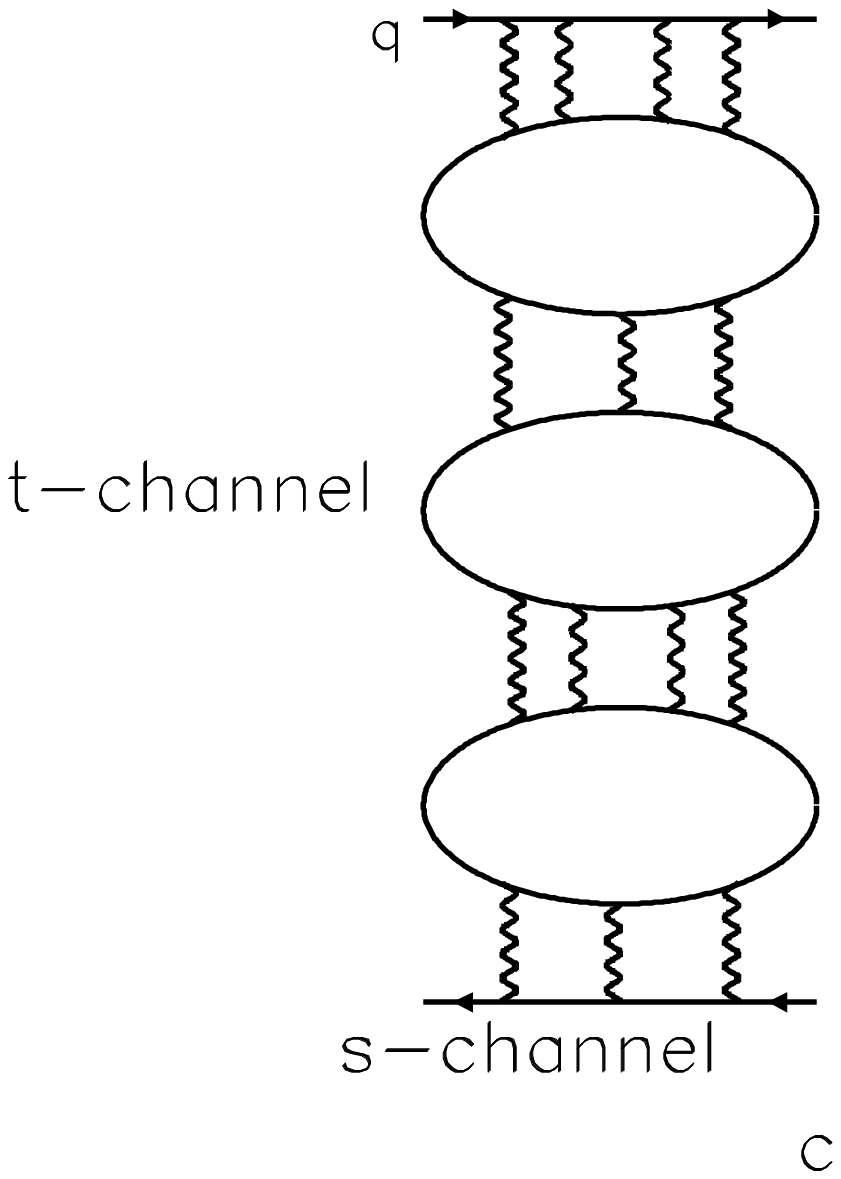,height=8cm}\hspace{1cm}
            \epsfig{file=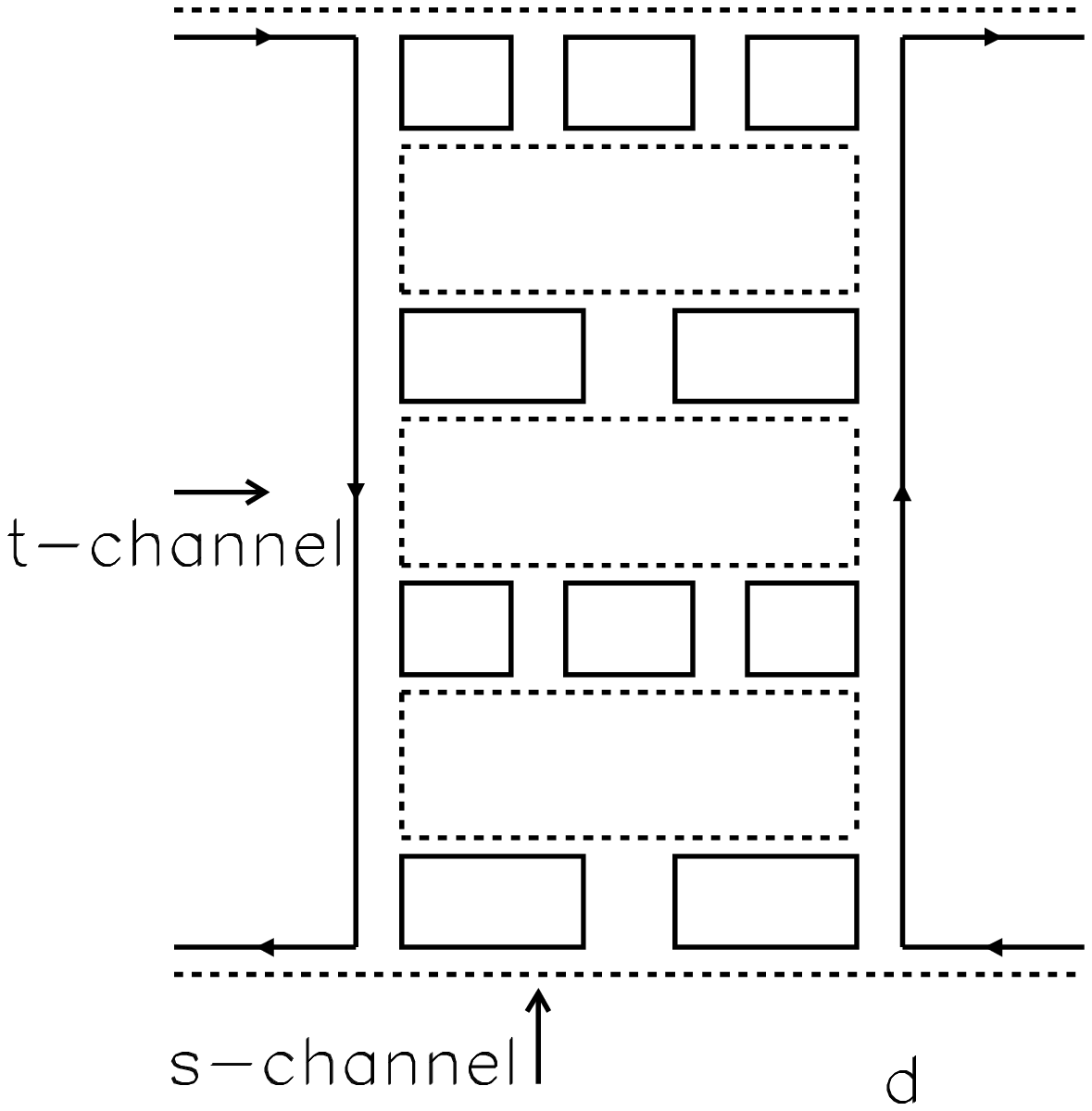,height=8cm}}
\caption{Diagrams for the block responsible for the confinement
interaction: a) transition $M^* \to hadrons$ with the chain of
quark-antiquark pairs; b) self-energy part the cutting of which gives
the diagram of Fig. 3a; c) the interaction block of the self-energy
diagram of Fig. 3b: a set of such diagrams yields the confinement
potential $V_8(r)$; d) the t'Hooft--Veneziano diagram for the block
of Fig. 3c: solid and dashed lines represent the propagation of colour
and flavour, respectively. }

\end{figure}

\end{document}